\documentclass[a4paper,twoside]{article}

\usepackage{listings}
\usepackage{xcolor}

\definecolor{codegreen}{rgb}{0,0.6,0}
\definecolor{codegray}{rgb}{0.5,0.5,0.5}
\definecolor{codepurple}{rgb}{0.58,0,0.82}
\definecolor{backcolour}{rgb}{0.95,0.95,0.92}

\lstdefinestyle{mystyle}{
    backgroundcolor=\color{backcolour},   
    commentstyle=\color{codegreen},
    keywordstyle=\color{magenta},
    numberstyle=\tiny\color{codegray},
    stringstyle=\color{codepurple},
    basicstyle=\ttfamily\tiny,
    breakatwhitespace=false,         
    breaklines=true,                 
    captionpos=b,                    
    keepspaces=true,                 
    numbers=left,                    
    numbersep=5pt,                  
    showspaces=false,                
    showstringspaces=false,
    showtabs=false,                  
    tabsize=2
}

\usepackage{epsfig}
\usepackage{enumitem}
\newcommand*\circled[1]{\tikz[baseline=(char.base)]{
            \node[shape=circle,draw,inner sep=.7pt] (char) {#1};}}
\newcommand{\specialcell}[2][l]{%
  \begin{tabular}[#1]{@{}l@{}}#2\end{tabular}}
\usepackage{tikz}
\usetikzlibrary{chains,positioning,arrows.meta, positioning, quotes,shapes}
\usepackage{listings}
\usepackage{hyperref}
\usepackage{subcaption}
\usepackage{calc}
\usepackage{amssymb}
\usepackage{amstext}
\usepackage{amsmath}
\usepackage{amsthm}
\usepackage{multicol}
\usepackage{pslatex}
\usepackage{apalike}
\usepackage[bottom]{footmisc}
\usepackage{csquotes}
\usepackage{listings}
\usepackage{caption}
\usepackage{SCITEPRESS}     

\begin{document}

\title{Identifying Personal Data Processing for Code Review}

\author{\authorname{Feiyang Tang\sup{1}, Bjarte M. {\O}stvold\sup{1}and Magiel Bruntink\sup{2}}
\affiliation{\sup{1}Norwegian Computing Center, Oslo, Norway}
\affiliation{\sup{2}Software Improvement Group, Amsterdam, The Netherlands}
\email{\{feiyang, bjarte\}@nr.no, m.bruntink@sig.eu}
}

\keywords{Data privacy protection, code review, static analysis}

\abstract{
Code review is a critical step in the software development life cycle, which assesses and boosts the code's effectiveness and correctness, pinpoints security issues, and raises its quality by adhering to best practices.
Due to the increased need for personal data protection motivated by legislation, code reviewers need to understand where personal data is located in software systems and how it is handled.
Although most recent work on code review focuses on security vulnerabilities, privacy-related techniques are not easy for code reviewers to implement, making their inclusion in the code review process challenging.
In this paper, we present ongoing work on a new approach to identifying personal data processing, enabling developers and code reviewers in drafting privacy analyses and complying with regulations such as the General Data Protection Regulation (GDPR).
}

\onecolumn \maketitle \normalsize 

\section{\uppercase{Introduction}}
\label{sec:introduction}
The General Data Protection Regulation (GDPR) lays the legal foundation for data protection in the EU and increases individual data protection rights throughout Europe. 
It also carries significant fines of up to 4\% of yearly worldwide revenue for businesses that do not comply with the legislation. 
Many IT system providers, especially software-producing firms, may need to alter their systems in order to comply with the GDPR. 
This is predicted to require significant effort~\cite{blume2016impact}. 
As a result, providing software engineers in the industry with effective and systematic ways to build data protection into software is an essential and beneficial study topic~\cite{lenhard2017literature}.
Organizations are pushing security to the software development life cycle, such as code review, to prevent software security vulnerabilities~\cite{braz2022software}.
Similarly, to comply with privacy-by-design and perform privacy analysis tasks, code reviewers would benefit from similar tools to those used for security to identify privacy-related patterns in software.

Developers address privacy concerns using data security terminology, and this vocabulary confines their notions of privacy to threats outside of the organization~\cite{hadar2018privacy}.
However, even though data security is the main prerequisite of data privacy, privacy protection in software is still very much different from traditional security-related vulnerabilities.
And according to Bambauer: ``security and privacy can and should be treated as distinct concerns''~\cite{bambauer2013privacy}.
Developers struggle to convert legal, ethical, and social privacy concerns into concrete technology and solutions~\cite{notario2015pripare}.

Assessing privacy involves not only finding personal data in the software but also evaluating compliance with the related processing.
GDPR defines as processing: ``any operation or set of operations which is performed on personal data or on sets of personal data, whether or not by automated means.''
The definition encompasses a vast range of actions performed on personal data, such as collecting, recording, organization, structuring, storage, adaption or modification, retrieval, transit, etc.
Privacy assessment tasks beg the question: How can we assist code reviewers and software developers in assessing personal data processing?
By identifying personal data and the relevant processing in the system, code reviewers can uncover interesting patterns and utilize them to redesign the system to be more privacy-friendly or perform privacy analysis.

In this paper, we present ongoing work on a novel approach designed to assist developers and code reviewers in identifying personal data processing, which can subsequently be used for privacy analysis.
This enables developers and code reviewers to assist organizations with a variety of important privacy-related tasks, such as completing a data protection impact assessment (DPIA) and creating a privacy policy.

\section{\uppercase{Related work}}
An essential step in the software development process, code reviewing incorporates both manual and/or automated reviews. 
The main goal of code reviews is to assess and boost the code's effectiveness and correctness, pinpoint security issues, and raise its quality by adhering to best practices~\cite{mcintosh2014impact}.
To automatically evaluate code, a variety of vulnerability detection tools have been built. 
They are also known as source code analyzers or static analysis tools, as they can analyze a program's code without having to execute it~\cite{mcgraw2008automated}. 

CodeQL\footnote{\url{https://codeql.github.com/}}, and Semgrep~\cite{semgrep}\footnote{\url{https://semgrep.dev/}} are two popular code review tools that utilize static analysis.
CodeQL treats code as if it were data, and issues are modeled as queries. 
Following the extraction of these queries from the code, they are executed against a database. 
The database is a directory containing data, a source reference for displaying query results, query results, and log files.
Semgrep matches grammatical patterns on parsed programs (represented as an Abstract Syntax Tree (AST)) instead of matching string or regular expression (regex) patterns on the program as a string.
Semgrep makes it considerably simpler to construct customized rules than CodeQL, which needs rules to be defined in QL, a declarative object-oriented query language.

There is relatively little published work that focuses on code reviews to identify privacy-related vulnerabilities, and it is problematic to translate current security knowledge to privacy, which we will explain in Section~\ref{Sec:challenges}.
There are studies on the identification of personal data that are valuable to our research.
Fugkeaw et al.~\cite{fugkeaw2021ap2i} proposed AP2I to enable organizations to identify and manage personal data in the local file system automatically. 
By monitoring network traffic, ReCon~\cite{ren2016recon} utilized machine learning to identify probable personal data breaches.
van der Plas et al.~\cite{van2022detecting} used CodeBERT, a RoBERT-like transformer model, to identify personal data in Git commits.

\section{\uppercase{Background and challenges}}\label{Sec:challenges}
Data privacy analysis is becoming as crucial as security vulnerability discovery and has brought a new dimension to the data security dilemma~\cite{bertino2016data}.
It is advantageous for code reviewers to be able to conduct a similar privacy analysis that they did for security.


The current state of the art is mostly focused on security analysis.
Although data security is a primary requirement for data privacy, the analysis domain and identification process are rather different~\cite{jain2016big}.
Simply adopting security mechanisms and mindsets to analyze privacy can be misguided, and even harmful~\cite{bambauer2013privacy}. 

Integration of recent studies on assessing software privacy during code review is challenging.
On the subject of program analysis, three well-known privacy analysis methods are available.
First, static analysis based on bytecode requires project compilation, whereas dynamic taint analysis requires project execution.
This is not practical nor efficient for code reviewers to implement.
A machine learning-based technique is similarly difficult to implement, as it requires a large and diverse training data set.
Obtaining and generating such data sets requires additional effort and could be outside the scope of code reviewers' capabilities.
Lastly, text analysis based on UI widgets is constrained for privacy by domain-specific UI attributes.
A financial web application that employs a model trained on an Android health mobile application is unlikely to benefit.
Code reviewers require an approach that is simple to deploy, efficient, and adaptable~\cite{buse2012information}.

Due to the complex nature of privacy and the fluidity of the definition of personal data, identifying the processing of personal data in the codebase presents challenges.

In the following paragraphs, we highlight the two most significant challenges related to the task in the context of code review.

\subsection{The Ambiguous Definition of Personal Data}
Article 4(1) in GDPR defines personal data as:

\begin{displayquote}
\textit{any information relating to an identified or identifiable natural person (‘data subject’); an identifiable natural person is one who can be identified, directly or indirectly, in particular by reference to an identifier such as a name, an identification number, location data, an online identifier or to one or more factors specific to the physical, physiological, genetic, mental, economic, cultural or social identity of that natural person.}
\end{displayquote}

The definition of personal data in the GDPR is so broad that almost any information may qualify as personal data if it refers to a specific individual, such as the fact that a person is wearing a red shirt~\cite{bervcivc2009identifying}. 
The definition is also semantically ambiguous. 

In contrast to the fact that certain data may be anonymous from the start (such as weather sensor data without any connection to real people), other data may initially be personal data but later be successfully altered to no longer have any connection to an identified or identifiable natural person. This emphasizes how flexible the categorization of personal data is~\cite{finck2020they}.

The same data point may be personal or non-personal depending on the context and may thus be covered by the regulation or not.
This implies that the categories of personal data in the software vary depending on the software and the processing underlying it.
For instance, health data such as blood pressure and medical records, for example, are sensitive for a health application, but location data is sensitive for navigation software.

Even if we accept that content-wise every item of information can be considered personal data if it can be related to an individual, the GDPR's definition is still rather vague structurally since it is not always clear what kind of structure every `record' of an individual must have to be considered personal data~\cite{voss2019personal}.

Due to the ambiguous nature of the definition of personal data in the relevant legislation, it is practically difficult for us to have a clear and fixed identifier to precisely locate personal data in code.

\subsection{What Counts as Sensitive Processing?}
Data subjects may agree to data processing for particular reasons.
This is the usual legal basis but only counts as one factor.
Processing may also be ``necessary for the performance of a contract to which the data subject is a party or in order to take steps at the request of the data subject prior to entering into a contract.''~\cite{voss2019personal}

Unfortunately, concerns that arise in principle about the relationship between contract and consent tend to be avoided in reality by disregarding consent requirements~\cite{pormeister2017informed}.

We cannot rely on existing privacy policies and written consent to uncover personal data processing in the codebase.
This requires us to consider all potential personal data processing in the codebase.
Later we will explain how we define and identify the relevant processing in software in Section~\ref{Sec:typeprocessing}.

\section{\uppercase{Approach}}
We present an approach to identify instances of personal data processing in the codebase and present them in a way that facilitates the code review.

The approach has three primary phases: pattern matching, labeling, and grouping of results.
As input, we take the codebase, which consists of source code files.
Then, a static analyzer will evaluate these source code files using our rules and patterns.
The code snippets discovered by the static analyzer are then labeled according to the various features they include.
Finally, we allow users to group the results by single or several labels, allowing a personalized exploration of the findings.

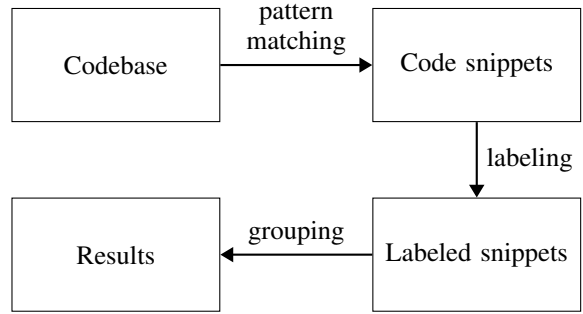
\begin{figure}[h!]
\centering
\begin{tikzpicture}[
    node distance=10mm and 20mm,
box/.style = {draw, minimum height=15mm, text width=25mm, align=center},
sy+/.style = {yshift= 2mm}, 
sy-/.style = {yshift=-2mm},
every edge quotes/.style = {align=center}
                        ]
\node (n1) [box]             {Codebase};                  
\node (n2) [box,right=of n1] {Code snippets};
\node (n3) [box,below=of n2] {Labeled snippets};      
\node (n4) [box,below=of n1] {Results};
\draw[thick,-Triangle]  
    (n1.east) to [above,"pattern\\matching"] (n2.west);
\draw[thick,-Triangle] 
    (n2.south) to [right,"labeling"] (n3.north);
\draw[thick,-Triangle]  
    (n3.west) to [above,"grouping"] (n4.east);
    
    \end{tikzpicture}
\caption{Approach}
\label{fig:flowchart}
\end{figure}

An illustration of our approach is shown in Figure~\ref{fig:flowchart}.

\subsection{Design Choices}
In the following paragraphs, we discuss our design choices for implementing the approach.

\subsubsection{Types of Findings}\label{Sec:typespd}
We want to have a basic default list of personal data that we want to locate, this is mostly personal identification and characteristics data, such as full name, email address, gender, sexual orientation, and age.
We call them fixed personal data.

According to different types of software, we customize default lists for them.
For example, for a banking/finance application, the list may contain bank account numbers, credit scores, and salary information.
This type of personal data is subject to context - the types of processing in specific software, which we named contextual personal data.

Depending on how we locate the mentioned personal data in the software, we can divide their occurrences in code into simply two types.

The first is in clear text. 
This includes all kinds of locations where personal data appear in clear text.
It is verbatim or direct personal data.
For example, a credit card number appears in an SQL query, or an email address falls into a log function.

The other type is more common and subtle, where personal data is stored in a variable or an object.
Depending on the different types of programming languages, the object types might vary from a local variable, a class instance, or a prototype.
This means we aim to find the code that processes this type of data.

\subsubsection{Types of Processing}\label{Sec:typeprocessing}
Simply locating every instance of personal data produces a large number of results.
Many of these do not directly help the code reviewer's work, which is to find meaningful processing.
We want to use a hybrid approach to cover as many as processing as possible.

Processing personal data represents a specific behavior.
This motivates our first approach: to use an action name tag to find relevant processing.
We adopted most of the verbs from Section 3 of DPV~\cite{pandit2019creating} \footnote{\url{https://w3c.github.io/dpv/dpv/}}. 
These vocabularies help us to find relevant processes in the software.

The second approach is the identification of external libraries.
We know that modern applications rely on various APIs to achieve different goals.
Therefore, obtaining a list of relevant APIs and detecting the existence of personal data that flows into them helps us find meaningful patterns.

\subsection{Pattern Matching}
The first step is to feed our codebase (consisting of source code files) to the static analyzer for pattern matching.
We chose Semgrep as our analyzer because of its user-friendly rules and rapid processing performance.
Depending on the different syntactic characteristics of personal data, as we discussed in Section~\ref{Sec:typespd}, we adopt a hybrid approach that combines two different types of analysis.

\begin{itemize}
    \item Match personal data in clear text using regular expression matching.
    \item Taint analysis to find flows in each file between a source (where personal data enters the analysis scope) and a sink (where personal data gets processed) that match our criteria.
\end{itemize}

Our personal data processing rules currently support Java, JavaScript, and TypeScript as our primary analysis domains.
However, our rules for identifying clear-text personal data apply to the vast majority of Semgrep-supported languages.

\subsubsection{Source and Sink}
Our prototype classifies the sources into nine separate categories.
As stated in Section~\ref{Sec:typespd}, we divide fixed personal data into four different categories: \textit{account}, \textit{contact}, \textit{national ID}, and \textit{personal ID}.
Included are five more contextual personal data categories, such as \textit{location}, \textit{health}, and \textit{financial} data.
In addition, we provide a template for identifying the processing of personal data and enable code reviewers and developers to submit additional personal data simply by entering the relevant keywords.
Then, corresponding rules will be automatically produced for future use.

Sinks are categorized into five main types. Three types of action: \textit{data manipulation} (\textbf{M}), \textit{data transportation} (\textbf{T}), and \textit{data creation/deletion} (\textbf{C/D}). Another two represent two special types: \textit{database} (\textbf{DB}) and \textit{encryption} (\textbf{E}).

A sink's name may contain a specific type of source. For example, \texttt{setLatitude(100,100)} \\ does not take any source into the method, but includes a source identifier \texttt{Latitude} and a sink identifier \texttt{set}, showing that it processes values directly as a source into a sink.
We call this special type of sink a source-specific sink.
When a source-specific sink invokes anything, we mark this source-specific sink as the new source but the caller of the source-specific sink as the new sink.
For example, in \texttt{gpsTracker.setLatitude(100,100)}, \texttt{setLatitude} becomes the new source and \texttt{gpsTracker} is the new sink.

Inspired by how Privado \footnote{\url{https://www.privado.ai}} uses regular expressions to identify GDPR-related data in Java applications, a sample Semgrep rule that matches the pattern of \textit{account} data source goes into a \textit{transportation} (\textbf{T}) sink is shown below in Figure~\ref{fig:rulesample}, followed by a sample code snippet detected in Figure~\ref{fig:samplecodesnippet}.

\begin{figure}[ht]
\includegraphics[width=\columnwidth]{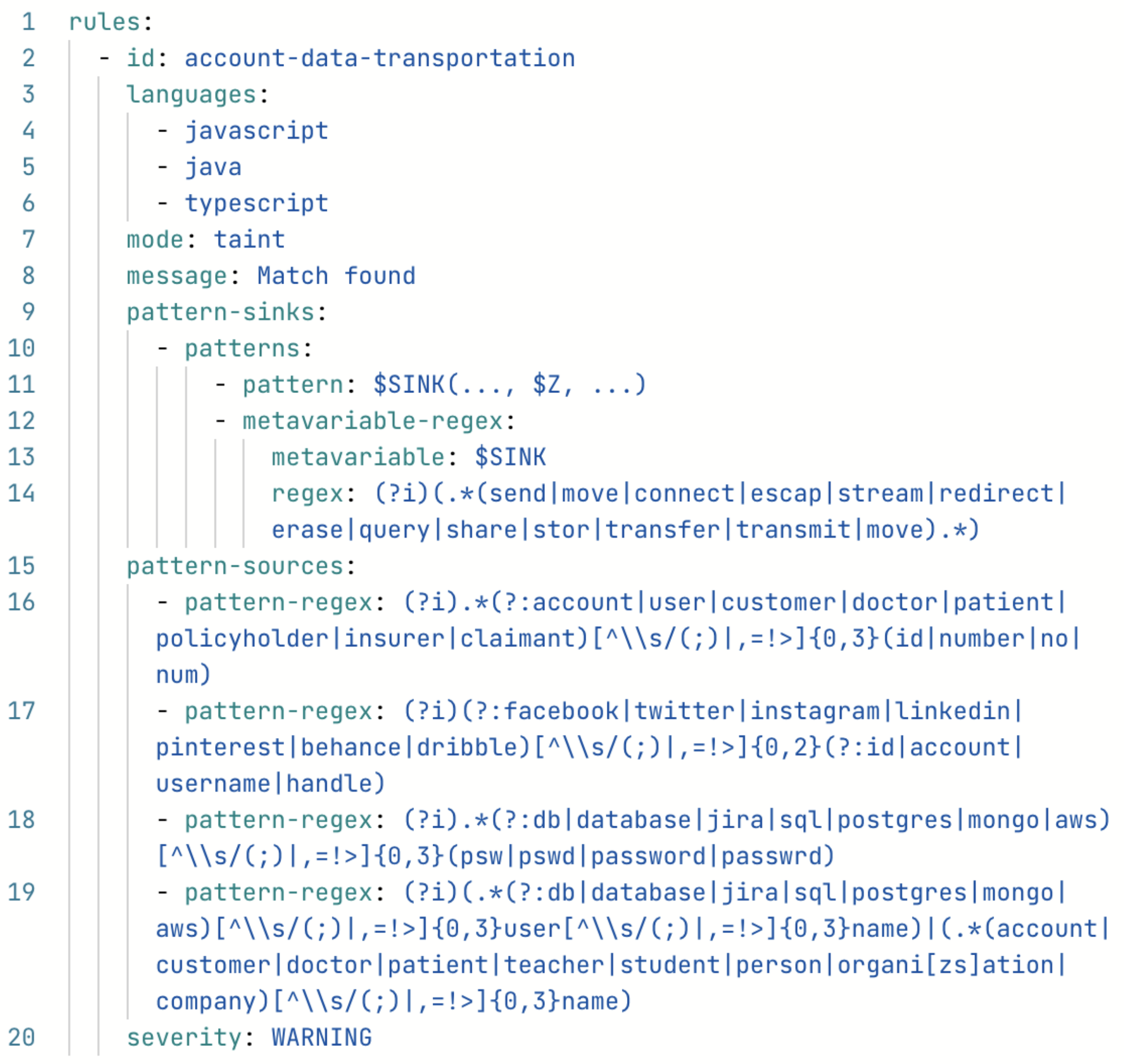}
\caption{Semgrep rule: find personal data flows from account data source to transportation sink}
\label{fig:rulesample}
\end{figure}

\begin{figure}[ht]
\includegraphics[width=\columnwidth]{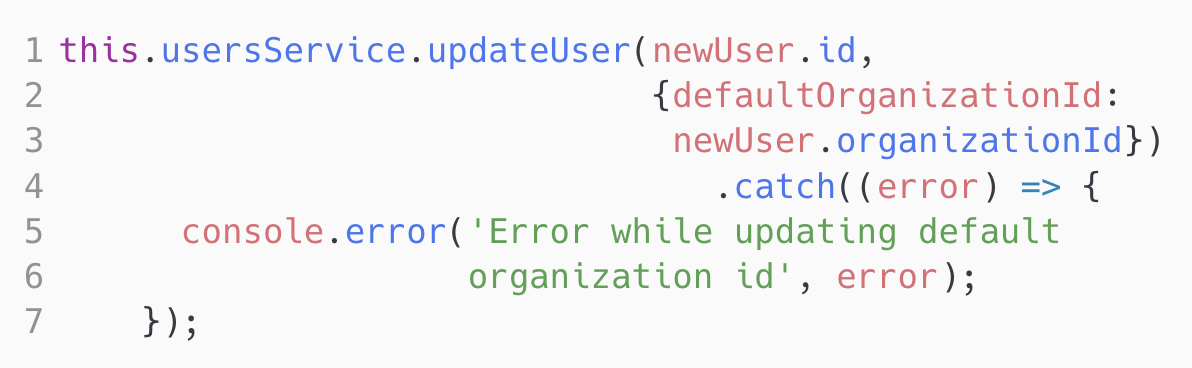}
\caption{Sample code snippet (from ToolJet) detected by Semgrep showing a flow from account personal data to a transportation sink.}
\label{fig:samplecodesnippet}
\end{figure}

\subsection{Labeling}
The identified findings from Semgrep are in the form of various lengths of code snippets (consisting of statements and expressions).
Each finding contains at least one detected sink and one source (or an object that received value from a source).
We abstract the structure of possible sources and sinks in each code snippet using the symbols below.

\begin{itemize}
    \item $O$ ranges over sources
    \item $I$ ranges over sinks
    \item $I^O$ ranges over source-specific sinks
\end{itemize}

We write $\bar{O}$ as shorthand for a possibly empty sequence $O_1,\cdots, O_n$.
Here the underscore $\_$ represents a placeholder for an expression that is insignificant in terms of privacy - it is neither a source nor sink nor contains a value from a source.

Below is a list of the common flow abstracts between sources and sinks that we observed in each code snippet.
Each abstract represents a typical flow, for example, \circled{1} to \circled{3} show that there are values passing through a sink to a source, from a non-privacy sensitive value (\circled{1}) or from another source (\circled{2}) or from innovating a sink inside another source object (\circled{3}).

\begin{multicols}{2}
    \begin{enumerate}[label=\protect\circled{\arabic*}]
        \item $O=\_.I(\_)$
        \item $O_2=\_.I(O_1,\_)$ 
        \item $O_2=\_.O_1.I(\_)$ 
        \item $\_=\_.O.I(\_)$ 
        \item $\_=\_.I(\bar{O},\_)$ 
        \item $\_.O.I(\_)$
        \item $\_.O.I(\_,\bar{O})$
        \item $\_.I^O(\_)$ 
        \item $\_.I^O(\_,\bar{O})$
        \item $\_.I(\bar{O},\_)$
    \end{enumerate}
\end{multicols}

For each identified code snippet, we label them with 22 labels (9 types of source, 5 types of sink, 5 types of source-specific sink, and 3 types of change in the sensitivity level), which are listed in Table~\ref{tab:notation}.
Besides the definition of source and sinks, we also introduce an important label: sensitivity.
The sensitivity level can increase, decrease, and stay the same in one identified code snippet.

\begin{table}[ht]
\centering
\begin{tabular}{ll}
 $\mathcal{O}$ & Nine types of source: $\{O^1, O^2, \ldots, O^9\}$ \\
 $\mathcal{I}$ & Five types of sink: $\{I^1, I^2, \ldots, I^5\}$ \\
 $\mathcal{I}^O$ & \specialcell{Five types of source-specific sink : \\ $\{{I}^{O^1}, {I}^{O^2}, \ldots, {I}^{O^5}\}$ }\\
 $\mathcal{S}$ & Sensitivity level change: $\{\texttt{up, down, equal}\}$
\end{tabular}
\caption{Labels to be assigned to each code snippet}
\label{tab:notation}
\end{table}

\paragraph{Sensitivity Level}
Not every result shares the same level of sensitivity regarding personal data processing.
After processing, the data from the source might remain at a similar sensitivity level, become more sensitive, or become less sensitive.

\begin{itemize}
    \item $\mathcal{S}=$ \texttt{up}: \circled{1}, \circled{4}, \circled{5}
    \item $\mathcal{S}=$ \texttt{equal}: \circled{2}, \circled{3}, \circled{6}, \circled{7}, \circled{8}, \circled{9}
    \item $\mathcal{S}=$ \texttt{down}: \circled{10}
\end{itemize}

\subsection{Result Presentation}
Johnson et al.~\cite{johnson2013don} pointed out that ``because the results
are dumped onto a code reviewer's screen with no distinct structure causing him to spend a lot of time trying to figure out what needs to be done''.
This indicates that developers and code reviewers may not benefit from ungrouped code snippets from static analysis tools if they are not presented in a sensible manner.

To tackle this issue, we present a two-phase technique to process the findings from Semgrep and present them to code reviewers in a smart way.

After each code snippet is labeled, we start to group them for presentation using their labels and other criteria.
Criteria for grouping include not only the labels but also other properties: 
\begin{itemize}
    \item neighboring results will be combined (same file and within a line number threshold);
    \item same or similar source/sink name; 
    \item same API usage (e.g. every code snippet that is related to the same API MongoDB).
\end{itemize}

\begin{figure}[ht]
\includegraphics[width=\columnwidth]{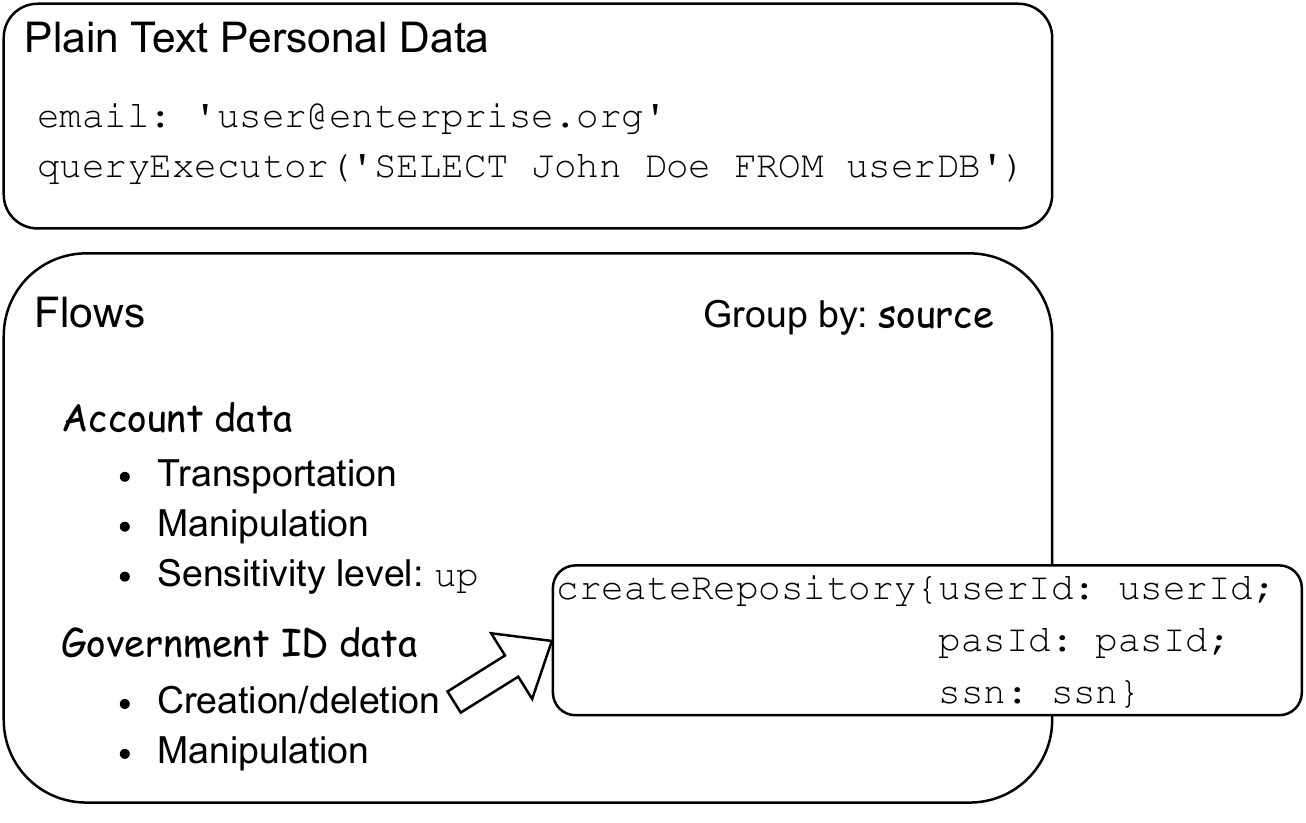}
\caption{Example presentation of the result. \textit{Personal data occurrences} is at the top and \textit{personal data processing code} is at the bottom.}
\label{fig:groupingeg}
\end{figure}

Figure~\ref{fig:groupingeg} following provides a straightforward illustration of how we present our results.
The results are presented in two separate sections: plain text results and flow results.
Users have the flexibility to select any label or label combination to filter the results.

\section{\uppercase{Demonstration}}
We created rules in Semgrep trying to capture as many useful findings for our analysis.
The software we analyzed here is ToolJet\footnote{\url{https://github.com/ToolJet/ToolJet}}, an open-source low-code framework for building React-based web applications.
ToolJet's implementation is mostly in JavaScript and TypeScript.
Users can build internal tools using ToolJet's prebuilt UI widgets to connect to data sources like databases, API endpoints, and external services.
This means ToolJet has many parts that process personal data, which makes it a good starting example.

Our Semgrep rules produce a total of 1,589 results from ToolJet's source code.
We manually reviewed each of the results and calculated the precision for each category.
If a single result can clearly demonstrate the processing of personal data, we consider it relevant and it could be beneficial for privacy code review.
Surprisingly, most false positives come from the personal data occurrence detector (with a precision of only 46.6\%), while most personal data processing results are relevant (with an average of 90.9\% precision for categories that have more than 50 code snippets identified).

Detailed statistics are listed in Tables~\ref{tab:result} and Table~\ref{tab:resultprecision}.


\begin{table}[ht]
\centering
\resizebox{\columnwidth}{!}{%
\begin{tabular}{cccccccc}
            & M  & T   & C/D & DB & E & L   \\ \hline\hline
Account     & 66 & 171 & 84  & 24 & - & 21  \\
Contact     & 89 & 175 & 36  & 3  & - & 3   \\
Personal ID & 56 & 133 & 41  & 7  & 1 & 4   \\
Online ID   & 6  & 26  & 1   & -  & - & 1   \\
Location    & 1  & 2   & -   & -  & - & -   \\
\end{tabular}
}
\caption{The code snippet count for each identified source and sink identified, `-' marks labels for which our approach detected no code snippet. Sink types are: \textit{data manipulation} (\textbf{M}), \textit{data transportation} (\textbf{T}), \textit{data creation/deletion} (\textbf{C/D}), \textit{database} (\textbf{DB}), \textit{encryption} (\textbf{E}) and \textit{log} (\textbf{L}).}
\label{tab:result}
\end{table}

\begin{table}[ht]
\centering
\resizebox{\columnwidth}{!}{%
\begin{tabular}{cccccccc}
            & M  & T   & C/D & DB & E & L   \\ \hline\hline
Account     & 90.9 & 90.6 & 95.2  & 91.67 & - & 95.2  \\
Contact     & 89.9 & 94.9 & 80.6  & *  & - & *   \\
Personal ID & 92.9 & 81.9 & 85.4  & * & * & *   \\
Online ID   & * & 84.6 & *   & -  & - & *   \\
Location    & *  & *  & -     & -  & - & -   \\
\end{tabular}
}
\caption{The precision of code snippet relevance (in \%) for each identified type of source and sink, `-' marks the labels for which our approach did not detect any code snippet, `*' marks the labels for which our approach detected less than 10 results. Sink types are: \textit{data manipulation} (\textbf{M}), \textit{data transportation} (\textbf{T}), \textit{data creation/deletion} (\textbf{C/D}), \textit{database} (\textbf{DB}), \textit{encryption} (\textbf{E}) and \textit{log} (\textbf{L}).}
\label{tab:resultprecision}
\end{table}

Figure~\ref{fig:samplecodesnippet} shows a simple interesting example of a grouped result showing how personal data \texttt{userId} is retrieved from a local repository in \textit{app\_users.service.ts} and then utilized to generate many data structures, such as the \texttt{app} object in \textit{app\_service.ts}.

\begin{figure}[ht]
\includegraphics[width=\columnwidth]{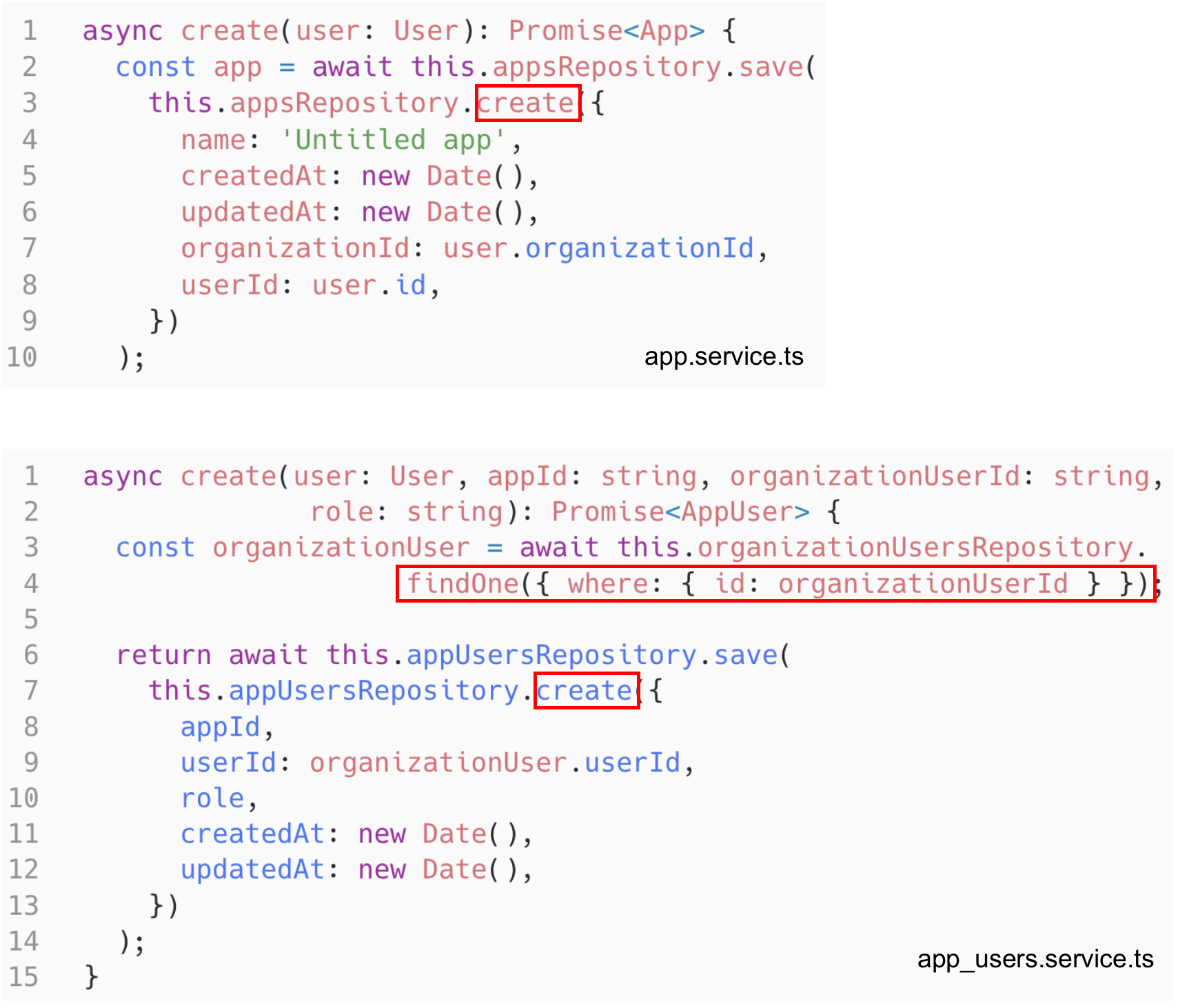}
\caption{Grouped example results showing how \texttt{organizationUserId} flows between functions.}
\label{fig:samplecodesnippet}
\end{figure}

\subsection{Future work}
Since our objective is to identify all relevant processing of personal data in source code, reducing false negatives is our next primary priority.
However, in our case, false positives are not a major concern.
Due to the subtlety of personal data processing, determining relevance without human assistance is particularly challenging.
Specifying the analysis to certain specific patterns would ease manual analysis.
This necessitates the implementation of a privacy taxonomy.
Using Ethyca's taxonomy~\cite{FidesLan43:online} as an example, we may modify our labels to match the technique with the taxonomy.

As an extension of this article, we propose an automated mapping of personal data in an unpublished (under review) manuscript~\cite{mapping} to assist developers and code reviewers in identifying privacy-related code.
The mapping based on static analysis automatically detects personal data and the code that processes it, and we offer semantics of personal data flows.

\section{\uppercase{Conclusions}}
\label{sec:conclusion}
This short paper presented ongoing work on a novel, customizable approach to identify personal data processing for code review.
This three-phase technique first uses Semgrep to match patterns in the code based on rules for sources and sinks, then associates code snippets generated from pattern matching with a set of behavioral labels, and finally groups results to reduce code reviewer workload.
Our demonstration shows the utility and feasibility of this method for gathering and presenting code snippets related to personal data processing from a codebase.

Along with the continued development of the approach architecture (refined rules for source and sink, more meaningful labels, and additional criteria for grouping), future work will focus on expanding the case study to include a larger set of open-source software from various domains and conducting a thorough user evaluation of the resulting platform.

\section*{\uppercase{Acknowledgements}}
This work is part of the Privacy Matters (PriMa) project. 
The PriMa project has received funding from European Union’s Horizon 2020 research and innovation program under the Marie Skłodowska-Curie grant agreement No. 860315.

\bibliographystyle{apalike}
{\small
\bibliography{paper}}

\end{document}